# Calculation of coercivity of magnetic nanostructures at finite temperatures

D. Suess, L. Breth, J. Lee, M. Fuger, C. Vogler, F. Bruckner, B. Bergmair, T. Huber, J. Fidler

*Vienna University of Technology, Wiedner Hauptstrasse 8-10, 1040 Vienna, Austria.*

T. Schrefl

*University of Applied Sciences, Matthias Corvinus-Straße 15, 3100 St. Poelten, Austria*

We report a finite temperature micromagnetic method (FTM) that allows for the calculation of the coercive field of arbitrary shaped magnetic nanostructures at time scales of nanoseconds to years. Instead of directly solving the Landau-Lifshitz-Gilbert equation, the coercive field is obtained without any free parameter by solving a non linear equation, which arises from the transition state theory. The method is applicable to magnetic structures where coercivity is determined by one thermally activated reversal or nucleation process. The method shows excellent agreement with experimentally obtained coercive fields of magnetic nanostructures and provides a deeper understanding of the mechanism of coercivity.

**I Introduction**

Since the original development of the theory of micromagnetism by W. F. Brown in 1940 [1][2], micromagnetic simulations became a standard tool for the design of various magnetic structures ranging from permanent magnets, magnetic recording devices, spintronic devices to magnetic logic elements. The success of micromagnetic simulations can be attributed to two developments (i) the tremendous increase of the CPU power [3] (ii) the fact that the lateral dimensions of most devices of interest shrunk from the micrometer regime to the nanometer regime. Hence, the number of the required cells of the discretization grid in order to describe the devices of interest decreased. The original micromagnetic theory as developed by Brown is a static theory that allows calculating equilibrium configuration of the magnetization. The implementation of the Landau –Lifshitz-Gilbert equation of motion allows describing the time evolution of the magnetization at zero temperature [4]. Finite temperature can be taken into account by adding a stochastic field to the effective field [5], [6], [7], [8],[9]. Solving the stochastic version of the Landau-Lifshitz-Gilbert equation successfully describes high frequency magnetic noise [10]. However, it is not practical feasible to describe the effect of temperature on magnetic properties such as the coercive field for measuring times in the regime of several seconds due to amount of computational time. A typical time step size $\Delta t$ of the stochastic version of the Landau-Lifshitz-Gilbert equation is in the range of $10^{-15}$s to $10^{-13}$s, which would result in $1/\Delta t$ time steps if a process of 1 second is simulated. Even for small systems simulations with such a large number of time steps are not performable on current computers. In order to describe the decay of the magnetization of magnetic recording media Monte Carlo Methods have been successfully applied [11],[12].

The dependence of the coercive field on the field rise time and temperature was studied in Ref. [15]. The reduction of the coercive field with increasing temperature agrees with VSM (vibrating sample magnetometer) measurements. This reduction can be attributed to thermal fluctuations. The fundamental understanding of coercivity as a function of both measurement time and temperature is of utmost importance for the future development of high density magnetic storage devices. A magnetic recording medium consists of many weakly coupled magnetic particles or grains. Traditionally one bit is formed by the ensemble of 20 or more grains. For the magnetic characterization of the data layer either the remanent coercivity or sweep rate dependent VSM measurements are used [16][19].

In a remanent coercivity measurement (i) the data layer is first saturated in a high positive field, (ii) then a constant reversed field is applied for a fixed time, (iii) the field is switched off and the remanent magnetization is measured. The procedure is repeated for progressively increased reversed fields. When the measured remanent magnetization is zero, half of the total magnetic volume is switched. The corresponding field $H_c$ is the remanent coercivity, which can be calculated as given by Sharrock [16] as,

$$H_c = H_0 \left( 1 - \left( \frac{k_B T}{\Delta E_0} \ln \left( \frac{f_0 t_0}{\ln(2)} \right) \right)^{1/n} \right) \qquad (1)$$

where $f_0$ is the attempt frequency, $k_B$ the Boltzman constant $\Delta E_0$ the energy barrier separating the two stable states at zero field, $H_0$ the zero temperature coercive field and $T$ the temperature. In order to derive Eq. (1) it is assumed that the energy barrier can be written as,

$$\Delta E_1 = \frac{\Delta E_0}{k_B T} \left( 1 - H_c / H_0 \right)^n \qquad , \qquad (2)$$

.

where $H_c$ is the external field that is applied in opposite direction with respect to the magnetization of the initial magnetic state. This field is kept constant and the time $\tau$ is measured which is required that the magnetization becomes zero. If this procedure is applied for different applied fields $H_c$ it allows to determine $\Delta E_0$ and $H_0$. Obviously, $H_0$ agrees with $H_c$ for $T$=0. The exponent $n$ is assumed to be a parameter in the range from 1 to 2 [21] . In principle Eq. (1) can be used to compute the remanent coercivity $H_c$ from the zero temperature coercive field $H_0$ and the energy barrier at zero field $\Delta E_0$. However, a detailed knowledge of the exponent $n$ and the attempt frequency $f_0$ is required for the computation of $H_c$ as function of waiting time and temperature. The exponent $n$ does not only depend on the angle between the external field and the easy axis direction but it also depends on the field strength [21],[15] . In addition the attempt frequency $f_0$ is unknown for reversal modes different from uniform rotation and depends on various factors such as field strength, field angle and temperature. Hence, generally the remanent coercive field at room temperature cannot be analytically obtained from the coercive field at zero temperature using Sharrock's equation. Hence, the aim of this paper is to present a numerical method to perform this task.

In contrast to the just discussed remanent coercivity measurement, in a sweep rate experiment the opposing external field not kept constant but it is linearly increased. At coercivity the measured magnetization crosses zero. Again at this point half of the magnetic volume in the magnetic recording medium is switched. Following the work of El-Hilo et al. for a linearly changing field the coercivity $H_c$ can be expresses as [31],

$$H_c = H_0 \left( 1 - \left( \frac{k_B T}{\Delta E_0} \ln \left( \frac{k_B T f_0 H_K}{2 \Delta E_0 R} \right) \right)^{1/n} \right) \qquad (3)$$

where the exponent in Eq. (3) is $n = 2$. Hence, this derivation is restricted to particles that reverse by uniform rotation with the external field applied perfectly parallel to the easy axis. Both methods (remanent coercivity and linear rising field) are equivalent in terms of the experimental characterization of the recording medium and lead to similar results with respect to switching field distribution of the particles and the thermal stability [16].

In this paper we describe a micromagnetic based theory that extends previous work to calculate the coercive field of nanostructures at finite temperature using energy barriers [17],[18] in a way that the coercive field is obtained from first principles [20] in the sense that the only input for the simulations are the measured material properties. All the other required properties are derived from these basic properties, such as the energy barrier and the attempt frequency which are calculated numerically using the transition state theory.

The presented theory goes beyond traditional micromagnetics as it includes temperate on a time scale of seconds and above. The paper is structured as follows: In section II, first a general numerical method is presented for calculating the coercive field at finite temperature that allows for an arbitrary external field as function of time. This method overcomes the before mentioned restrictions and can be used to describe sweep rate experiments. In the following of section II this general method is used to derive the special case of the constant field finite temperature micromagnetics (corresponding to remanent coercivity measurements). In this section also a short review on the numerical calculation of the attempt frequency is given which is used in the simulations. In section III the developed numerical methods are applied to two independent experiments of magnetic granular films. The importance of calculation at finite temperature is summarized in section IV.

**II Methods**

**(a) Coercive field for arbitrary field  as function of time**

In the following we will describe a numerical method for the computation of the coercivity of magnetic structures at finite temperature. The proposed method is applicable to systems where the coercivity is determined by a thermally activated reversal process in some selected region of the structure. The critical field when this reversal happens will be calculated. For example the method can be used to calculate the critical field at finite temperature of permanent magnets. In particular the field is calculated, when a nucleation in a grain of the hard magnet is formed.

 Another example are recording media which are composed of weakly coupled grains so that reversal occurs grain by grain.  The coercive field of a representative grain in a media can be calculated.

In the following we will use in the derivation the language suitable for ensemble of grains. However, instead of having an ensemble of grains, the method is also suitable to describe the average coercive field of one grain (e.g. hard magnet) if the measurement is repeated several times.

Let us take one grain out of an ensemble of grains. At zero temperature the coercive field of the grain is given if the energy barrier between the state pointing up and the state pointing down becomes zero.  At finite temperature thermal fluctuations help to reverse the particle. At the switching field the energy barrier is not zero.  The Arrhenius-Néel law gives the average lifetime of one grain as a function of the energy barrier

$$\tau = \frac{1}{f_0} e^{\Delta E_i(H)/k_B T} \qquad\qquad (4)$$

.

In Eq. (4)  Néel assumed that the energy barrier between the two equilibrium states is larger than the thermal energy $k_B T$ [22],[23].  Experimentally the Arrhenius-Néel (Néel-Braun) model was supported by micro-SQUID measurements on individual Co nanoparticles using waiting time, switching field and telegraph noise measurements [24].

Let us further assume that the grains can be described by a two level system. One level corresponds to the state with magnetization up, the other level corresponds to the state with magnetization down. The occupation probabilities of the two energy levels $P_1$ and $P_2$ satisfy the normalization condition $P_1 + P_2 = 1$ and the master equation

$$\frac{dP_1}{dt} = -w_{12}(t)P_1 + w_{21}(t)P_1 \qquad (5)$$

where $w_{12}$ is the transition rate from the up state to the down state and $w_{21}$ the transition rate from the down to the up state. $w_{12}$ is the inverse of the average lifetime of the up state, $w_{12} = \frac{1}{\tau}$.

Generally $w_{12}$ and $w_{21}$ will depend on time, since a time varying external field changes alters the lifetime of the particle.

For sufficiently large downward fields the up state has a much larger energy than down state. In this limit $w_{21}$ is much smaller than $w_{12}$ and can be set to zero. Under this assumption it follows,

$$\frac{dP_1}{dt} = -w_{12}(t)P_1 \qquad , \qquad (6)$$

which has the solution

$$P_1(t_{final}) = P_1(0)\exp\left(\int_0^{t_{final}} -w_{12}(t)dt\right). \qquad (7)$$

In the following we assume that the magnetic structure is saturated. Hence, $P_1(0) = 1$.

For the numerical calculation of Eq. (7), we divide the time into discrete time intervals $\Delta t$ with $t_i = i \, \Delta t$,

$$P(H_c) = \exp\left(\sum_{i=0}^{i=t_{final}/\Delta t} -\frac{1}{\tau_{t_i}}\Delta t\right) = \prod_{i=0}^{i=t_{final}/dt} P_{t_i} \qquad , \qquad (8)$$

where $P_{t_i} = \exp\left(-\Delta t / \tau_{t_i}\right)$ is the probability that a particle does not switch at time $t_i$ within the time interval $\Delta t$.

Coercivity is reached when half of the particles are switched. Therefore the coercive field is determined by the condition that

$$P(H_c) = 0.5 \qquad\qquad (9)$$

This method (arbitrary field FTM) allows for the calculation of the time $t_{final}$ and hence allows determining the field at this time, which is the coercive field.

This flexible method, which is called in the following arbitary field finite temperature micromagnetics (FTM) method, allows for the calculation of $\mu_0 H_c$ the coercive field at room temperature of complicated microstructures such as patterned elements [13], permanent magnets [14] or exchange spring media [15]. Solving Eq. (9) requires the value for the attempt frequeny $f_0$. The method used to compute $f_0$ is described in the next section, which allow taking in account the dependence of the attempt frequency as function of the damping constant, material parameters but also on the external field $H$.

**(b) Coercive field at constant field**

If after the quick application the external field does not change as function of time Eq. (7) can be further simplified to allow an even more efficient numerical calculation of the coercive field at finite temperature. From Eq. (7) together with the condition of coercivity (Eq. (9) ) it follows[1],

$$\exp\left(-f_0 t_{final} e^{-\Delta E_1 / k_B T}\right) = 0.5 \quad . \tag{10}$$

Solving Eq. (10) with respect to the energy barrier, it follows

$$\Delta E_1 = k_B T \ln\left(\frac{f_0 t_{final}}{\ln(2)}\right) \tag{11}$$

Therefore the problem of calculating the remanent coercivity is reduced to the problem of calculating the field strength that leads to an energy barrier $\Delta E_1$. This problem is solved iteratively using the Newton method. The following nonlinear equation has to be solved,

$$G(H_c) = \Delta E(H_c) - \Delta E_1 = 0 \tag{12}$$

The unknown variable is the external field strength $H_c$. The energy barrier as function of field $\Delta E(H_c)$ is calculated using the nudged elastic band method for every external field strength $H_c$ [26]. One prerequisite of the method is that $G(H_c) = 0$ has to exist. From nucleation theory it is known that this prerequisite is fulfilled for the single droplet nucleation regime but not in the Kolmogorov-Mehl-Johnson-Avrami regime [25].

**(c) Calculation of the attempt frequency**

---

[1]For completeness we want to note that solving Eq. (7) together with Eq. (9) with respect to $H_c$, directly leads to Sharrock's equation (Eq. (1) ).

The calculation of the constant field and arbitrary field FTM method require the calculation of the attempt frequency $f_0$ in each iteration which depends on various factors such as external field, magnetic microstructure and temperature. For the numerical calculation of the attempt frequency we follow the approach of Langer of the transition state theory [27]. The attempt frequency can be written as:

$$f_0 = \frac{\lambda_+}{2\pi} \Omega_0 \qquad . \qquad (13)$$

where $\lambda_+$ denotes for the dynamical prefactor and $\Omega_0$ is the ratio of the well and saddle angular frequencies. For the calculation of the attempt frequency the total Gibbs free energy of the system as well the linearized equation of motion around the saddle point have to be expressed in its canonical variables. We use the finite element method to discretize the continuous magnetization on $N$ - node points. On each of the node point $i$ the canonical variables are given by $p_i = V_i J_s \cos(\theta_i)$ and $q_i = \phi_i$, where $V_i$ is the corresponding volume, which is assigned to the node point $i$. The angle $\theta_i$ and $\phi_i$ describe the direction of the magnetization. In a next step the Hessian matrix of the total Gibbs free energy which respect to $p_i$ und $q_i$ has to be calculated. The ratio of the determinates of the Hessian matrix evaluated at saddle point and at the minimum of the metastable state of total Gibbs free energy gives,

$$\Omega_0 = \sqrt{\frac{\det\left(H_{stat}\right)\big|_{\min}}{-\det\left(H_{stat}\right)\big|_{sp}}} \qquad . \qquad (14)$$

As a consequence the coercivity at finite temperature calculated with the method presented above depends also on the metastable state.

The dynamical prefactor $\lambda_+$ accounts for the equation of motion of the system. It can be derived from an eigenvalue problem of the linearized Landau-Lifshitz-Gilbert equation around the saddle. Details are given in Ref [28], [29], [30].

**III Results**

**(a) Angular dependence of the coercive field**

In order to test the arbitary field FTM method we compare the results with experiments. In the micromagnetic model we calculated the coercive field of only one grain of the perpendicular recording media. At the coercive field of a particle ensemble 50% of the magnetic volume points up the other 50% is pointing down. Hence, the average stray field at $\mu_0 H_c$ can be approximated to be zero [15]. Details of the experimental measurements of the CoCrPtO perpendicular recording media can be found in Ref [32]. The granular film has a thickness of 11.3 nm. The average grain size of the granular film is 7.5 nm. From the measured anisotropy field an average anisotropy constant and magnetic polarization of $K_{1,average} = 0.3$ MJ/m³ and $J_{s,average} = 0.5$ T is obtained, respectively. For the particular film the fraction of the total film volume ($V_{total}$) to the volume of the magnetic grains ($V_{grain}$), without a 1 nm thick non magnetic grain boundary region is,

$$r = \frac{V_{grain}}{V_{total}} = 0.77 \ .$$

Using this ratio we can derive from the measured average anisotropy and polarization, the anisotropy constant of the grain $K_{1,grain} = K_{1,average}/r = 0.386$ MJ/m³ and magnetic polarization $J_{s,grain} = J_{s,average}/r = 0.644$ T, respectively.

Since the simulation of the coercive field is simplified to the calculation of $H_c$ of one representative grain a detailed knowledge of the grain size distribution is required. We obtain the grain size distribution $f(D)$ from the experimental data as,

$$f(D) = A_0 \exp\left(-\frac{1}{2}\left(\frac{D-D_0}{\sigma}\right)^2\right), \text{ with } D_0 = 7.5 \text{ nm and } \sigma = 0.99 \ .$$

In the following we assume that due to thermal fluctuation first the small grains are going to be reversed due to a smaller thermal stability. Under this assumption we can calculate an effective grain diameter $D_{eff}$. If $V_{tot}$ is the total magnetic volume of the particle ensemble, the volume of all grains with $D < D_{eff}$ equals $V_{tot}/2$. The volume of the grains is proportional to $V \sim D^2$ under the assumption of a constant film thickness. Using the grain size distribution $f(D)$ we can compute $D_{eff}$ from the following equation

$$\int_0^{D_{eff}} D^2 f(D) dD = \int_{D_{eff}}^{\infty} D^2 f(D) dD \qquad (15)$$

giving $D_{eff} = 7.76 nm$ for CoCrPtO media. In the following we use $D_{eff}$ as the representative grain size diameter for the calculation of the energy barriers. In the calculation grains with hexagonal basal planes are used are used as shown in Fig 1. The diameter Deff of the hexagon is defined as ,

$$(D_{eff} / 2)^2 \pi = A \qquad (16)$$

where $A$ is the area of the hexagon.

VSM hysteresis loops were performed for different angles θ between the external field and the easy axis. The field sweep rate is 0.5 T/s. The measured angular dependence of the coercive field is shown by the (black) solid (triangles down) line in FIG. 1. The dashed (red) line with circles shows the calculated values of the coercive field using the measured material parameters as discussed before. The obtained values of the coercive field are much larger than the experimental ones because the SW-model does not account for thermal fluctuations. Besides the significant disagreement between the absolute values of the coercive field of the –SW- model with the experimental data, an additional mismatch is the form of the angular dependence. FIG. 1 also shows SW angle dependence of the coercive field scaled so that the values match at 45 degrees. Fig. 1 shows that there is a significant disagreement for small angles α as ( (red dashed lines and red squares). The decrease of the coercive field as the angle θ is changed from θ = 0° to θ = 45° is smaller in the experiment than predicted by the SW-theory. According to the SW- model the coercive field shows a decrease by 50% as the field is changed from a direction parallel to the easy axis to θ = 45°. The experimental data however show that the coercive field $\mu_0 H_c$ at θ = 45° is reduced only by about 32%. Gao et al. has suggested that exchange coupled misaligned grains lead to a shallower angular dependence of the coercive field as compared to the SW-theory [33]. In this work we show that the experimentally obtained angular dependence of the coercive field is perfectly reproduced solely by taking into account thermal fluctuations. In principle the discrepancy between the angular dependence predicted by the SW- model and the experimental values can be understood by investigating Sharrock's equation.

In order to understand the angular dependence of the coercive field, we perform micromagnetic simulations using the arbitrary field FTM. The (blue) diamonds in FIG. 1 show the coercive fields for a temperature of $T = 300$K and a field rise time of R = 0.5 T/s, which is the same rise time as in the experiment using the arbitrary field FTM. An almost perfect agreement with the experimental data is obtained with regards to the absolute values as well as the angular dependence.

In order to point out the importance to use $K_1 = K_{1,grain}$, $J_s = J_{s,grain}$ and not to use the average measured values $K_{1,average}$, $J_{s,average}$ for the material parameter of a grain, we plot in FIG. 1 the coercive field if $K_1 = K_{1,average} = 0.3 MJ / m^3$ and $J_s = J_{s,average} = 0.5T$ is assumed (black dotted line, black triangles up). Due to the smaller anisotropy constant the particle is less thermally stable, which leads to a significant reduction of the coercive field.

We also compare the results of the arbitrary field FTM with the results of the constant field FTM using the approximation of El-Hilo et al. The maximum difference of the coercive field obtained from Eq. (3) (El-Hilo) and the numerically obtained coercive field is 2%. This good agreement indicates that (i) the investigated sample is sufficient small that reversal mode can be well approximated with the SW – Theory and (ii) that the assumption that $n = 2$ does not introduce significant errors in the calculation of $H_c$. Of course, for more complicated structures which do not reverse via homogenous rotation the numerical method as developed in section II is essential for reliable predictions as shown in the next section.

FIG. 2 shows the attempt frequency which is calculated at each time step of Eq. (8) to determine the coercive field at finite temperature. The attempt frequency is plotted as a function of the angle for the external field that equals the coercive field. The attempt frequency increases from 753 GHz to 823 GHz with increasing angle. The angular dependence of the attempt frequency is mainly determined by the angular dependence of $\lambda_+$ as shown in FIG. 2.

**(b) Thickness dependence of the coercive field**

As a second example we apply the constant field FTM method for the calculation of the coercive field of a perpendicular recording media as a function of the layer thickness as shown in FIG. 3. The experimental data is extracted from the work of Wu et al. [34]. In the paper the average grain diameter was given as 7 nm. The coercive field of a 41 nm thick film at $T = 2$ K was measured to be $\mu_0 H_c = 1.2$ T. This information was used to extract the anisotropy constant for the simulations ( $K_{1,\text{grain}} = 0.345$ MJ/m³ , $J_{s,\text{grain}} = 0.64$ T). The external field was applied 2° off the easy axis. We calculated the coercive field as a function of the layer thickness for a temperature of $T = 300$ K. In the simulations, the field waiting time is $t_0 = 0.027s$, which corresponds for this system to a rise time of 0.5 T/s. Due to thermal fluctuations the coercive field is drastically decreased for smaller layer thickness. For grains thinner than $l_{crit} = 4\sqrt{\dfrac{A}{K}}$ the magnetic reversal mode is homogenous rotation. Then the energy barrier increases linearly with particle volume (and the layer thickness). Exceeding the critical thickness $l_{crit}$ a domain wall is formed during magnetic reversal leading to an energy barrier which becomes independent on the particle length. As a consequence for films thicker than $l_{crit}$ the coercive field at finite temperature also becomes independent on the particle length. Although the reversal mode changes as the layer thickness is increased the attempt frequency only increases by about 20% as the reversal model changes from coherent rotation to nucleation.

Fig. 3 shows that the dependence of the coercive field as a function of the film thickness can be explained very well with the presented model. A comparison of the coercive field obtained from Eq. (1) with the experimental data is given in Ref [28]. In particular for layer thickness $l > l_{crit}$ Eq. (1) does not agree with the experimental results as shown in Ref [28]. This emphasizes the importance of the developed method of this paper.

**IV Summary**

To conclude, the developed FTM method goes beyond standard micromagnetic simulations, where the coercive field is generally obtained if the energy barrier, between two stable states vanishes. The method allows for the calculation of the coercive field of arbitrary shaped magnetic nanostructures at finite temperature at time scales of nanoseconds to years if coercivity is determined by one significant thermally activated reversal process. Since the proposed method relies on the transition state theory it requires that at the coercive field the energy barrier $\Delta E_1 >> k_B T$. This is fulfilled if the investigated magnetic structure is thermally stable and the external field is applied with a sufficient low field rise rime. Furthermore, it is assumed that the thermally activated process can be described by a two level system, where due to the external field the switching from the initial state to the reversed state is much more probably than switching back again.

The proposed method is essential for the interpretation of magnetic measurements of various magnetic structures ranging from advanced storage concepts [13] to permanent magnets [14], where state of the art simulations are done at zero temperature and hence may lead to incorrect conclusions.

We discovered that fundamental magnetic properties such as the angular dependence of the coercive field of magnetic nanostructures significantly changes if thermal fluctuations are taken into account.

Helpful discussion with Hong-Sik Jung and the financial support of the WWTF project MA09-029 and Austrian Science Fund (FWF:) SFB ViCoM ( F4112-N13 ) is acknowledged.

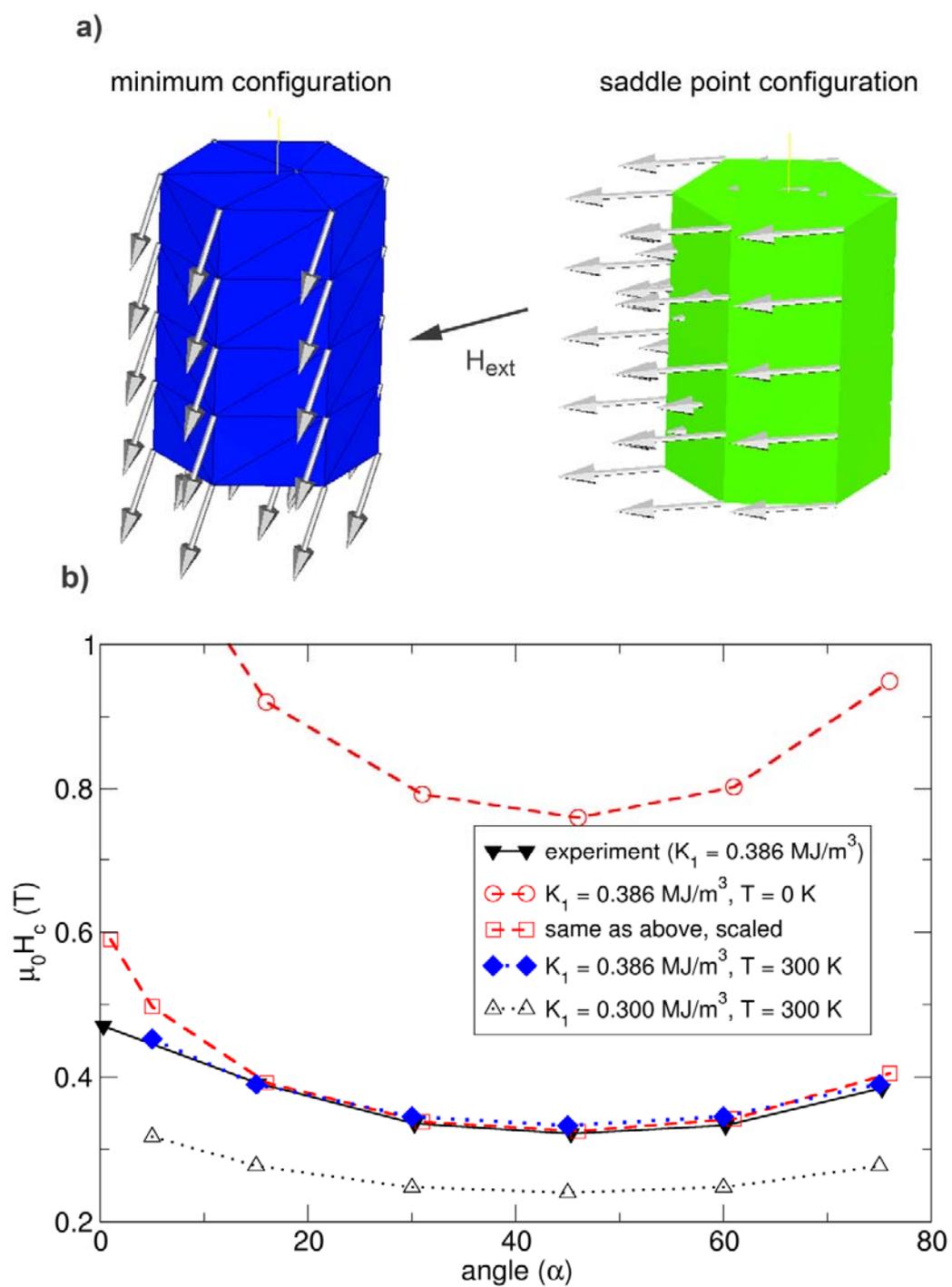

**a)**

minimum configuration                     saddle point configuration

$H_{ext}$

**b)**

FIG. 1 (color online): (a) Geometry and finite element mesh of the model of a granular grain. The initial magnetization and the saddle point configuration at the coercive field for a field applied 75° off the long axis of the particle is shown. These states are used for the calculation of the attempt frequency. (b) Angular dependence of the coercive field ($\mu_0 H_c$) of a CoCrPtO granular medium. (black solid line, triangles down) experimental obtained  values extracted from Ref. [32].  (red dashed line, red circles) calculations at zero temperature, with the material parameters extracted from the experiment. (red dashed line, red squares) calculations at zero temperature coercive field, where the anisotropy constant is scaled to match the coercive field at $\alpha = 45°$.  (blue diamonds) calculations at room  temperature without any free parameter with the measured material parameters using the arbitrary field FTM method.  (black dotted line, black triangles up) coercive field, if the measured values of the anisotropy constant and magnetic polarization are directly assumed to be the proper material parameters for the grain (

$K_1 = 0.3 MJ / m^3$ and $J_s = 0.5 T$ )

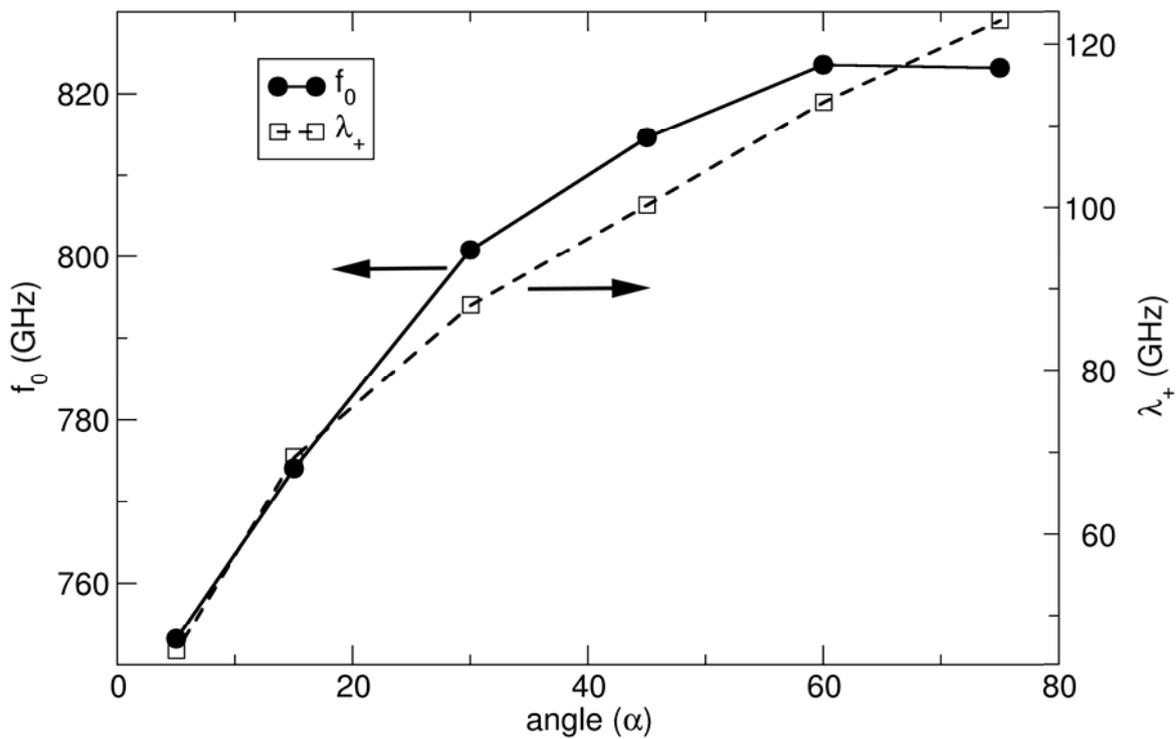

FIG. 2 : Self consistently calculated attempt frequency ($f_0$) using the FTM method for the calculation of the coercive field. The attempt frequency is determined by $\lambda_+$ which is also plotted.

saddle point configurations:

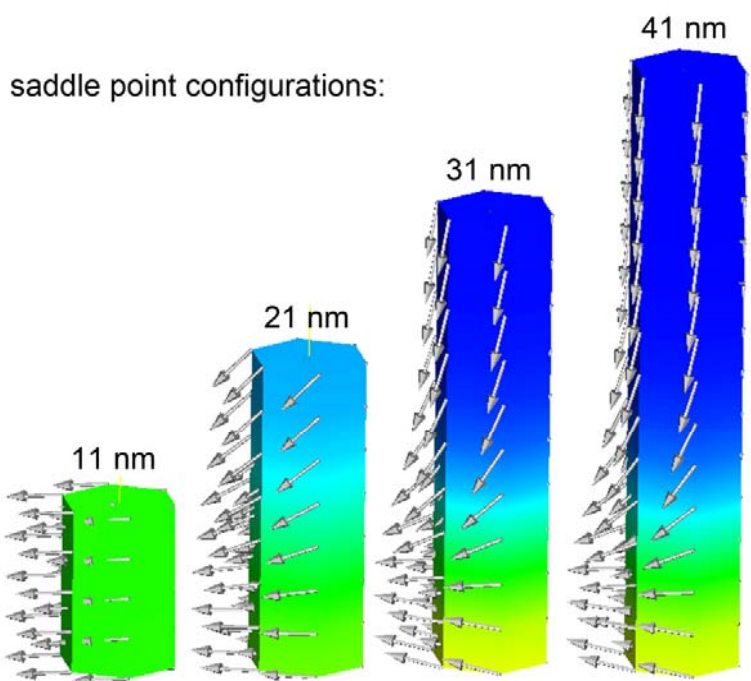

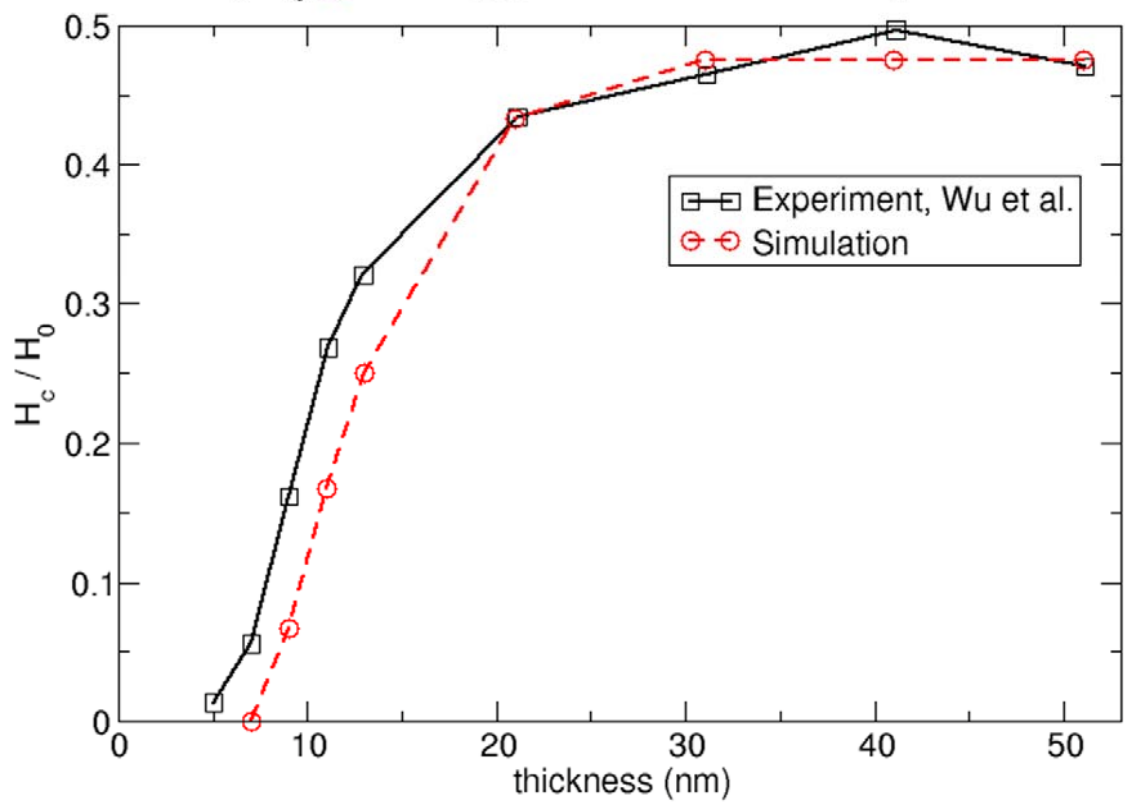

FIG. 3 (color online): Comparison of the calculated (constant field FTM) and measured coercive field of a perpendicular recording media for different film thicknesses at T = 300 K. The experimental data is extracted from the work of Wu et al. [34]. In the simulation a grain with a hexagon as basal plane with with $D$ = 7 nm is assumed. $J_{s,grain}$ = 0.64 T, $K_{1,grain}$ = 0.345 MJ/m³. The magnetic configuration at the coercive field at the saddle point is shown for four different grain thicknesses.